\documentclass[                     
               preprintnumbers,     
               aps,                 
               prd,                 
               nofootinbib,         
               floats,floatfix      
               ]{revtex4}

\usepackage{amssymb,graphicx}
\begin{document}

\title{Absorbing boundary conditions for Einstein's field equations}

\date{\today}
\author{Olivier Sarbach}
\affiliation{Instituto de F\a'{\i}sica y Matem\a'aticas,
Universidad Michoacana de San Nicol\a'as de Hidalgo, Edificio C-3, 
Cd. Universitaria. C. P. 58040 Morelia, Michoac\a'an, M\a'exico}

\begin{abstract}
A common approach for the numerical simulation of wave propagation on
a spatially unbounded domain is to truncate the domain via an
artificial boundary, thus forming a finite computational domain with
an outer boundary. Absorbing boundary conditions must then be
specified at the boundary such that the resulting initial-boundary
value problem is well posed and such that the amount of spurious
reflection is minimized. In this article, we review recent results on
the construction of absorbing boundary conditions in General
Relativity and their application to numerical relativity.
\end{abstract}


\maketitle

\section{Introduction}
\label{Sect:Intro}

To numerically simulate the evolution of hyperbolic partial
differential equations on a spatially unbounded domain, one usually
replaces the unbounded domain with a finite, compact domain $\Omega$
with artificial outer boundary $\partial\Omega$. Boundary conditions
on $\partial\Omega$ must then be specified in order to obtain a unique
Cauchy evolution. These conditions should be formulated so that they
form a well posed initial boundary value problem (IBVP) and, ideally,
are completely transparent to the physical problem on the unbounded
domain. In practice, complete transparency cannot be achieved easily
and therefore, the boundaries introduce spurious reflections into the
computational domain $\Omega$. The idea then is to develop what is
called absorbing boundary conditions which form a well posed IBVP and
insure that the amount of spurious reflection from $\partial\Omega$ is
{\em as small as possible}.

There has been a substantial amount of work on the construction of
absorbing (also called non-reflecting in the literature) boundary
conditions for wave problems in acoustics, electromagnetism,
meteorology, and solid geophysics (see \cite{dG91} for a review). One
approach is based on a hierarchy of {\em local} boundary conditions
\cite{bEaM77,aBeT80,rH86} with increasing order of accuracy. Although
higher order local boundary conditions usually involve solving a high
order differential equation at the boundary, the problem can be dealt
with by introducing auxiliary variables at the boundary surface
\cite{dG01,dGbN03}. A different approach is based on fast converging
series expansions of {\em exact nonlocal} boundary conditions (see
\cite{bAlGtH2000} and references therein).

Constructing absorbing outer boundary conditions in General Relativity
is difficult. First of all, Einstein's field equations determine the
evolution of the metric tensor, so one does not know the geometrical
structure of the spacetime before actually solving the IBVP. Hence, it
is not clear {\em a priori} how the geometry of the outer boundary
evolves. One could turn the argument around to guide the choice of
boundary conditions so that they fix the embedding of the boundary
surface into the resulting spacetime. Second, in the Cauchy
formulation of Einstein's field equations, one usually encounters
constraint fields which propagate {\em across} the boundary into the
computational domain. This is in contrast to the standard Cauchy
formulation of Maxwell's equations, where the evolution equations
imply that the constraint variables (namely, the divergence of the
electric and magnetic fields) are constant in time. As a consequence,
in the Einstein case, constraint-preserving boundary conditions need
to be specified at $\partial\Omega$ such that constraint violations
are not introduced into the computational domain. Finally, due to the
nonlinear nature of the theory and its diffeomorphism invariance, it
is difficult to define precisely what is meant by in- and outgoing
radiation in General Relativity. Therefore, it is not even clear how
to quantify the amount of spurious gravitational reflection from the
boundary.  These issues all contribute to the challenge of developing
accurate absorbing boundary conditions for Einstein's field equations.

In this article, we briefly review some recent results on the
construction of absorbing outer boundaries in General Relativity. We
start with a discussion of absorbing outer boundaries for the wave
equation on Minkowski and Schwarzschild spacetimes in
Sect. \ref{Sect:ABCWaveEq} where we review results from the literature
and derive asymptotic outgoing wave solutions on weakly curved
spacetimes. Recent attempts to generalize these results to Einstein's
field equations are mentioned in Sect. \ref{Sect:ABCEinsteinEq}. In
particular, we discuss constraint-preserving boundary conditions,
boundary conditions designed to minimize spurious reflections of
gravitational waves, recent results on well posed initial-boundary
value formulations and recent applications to numerical
relativity. Concluding remarks are drawn in
Sect. \ref{Sect:Conclusions}.

There exist other approaches for dealing with gravitational wave
propagation on an infinite domain which are not discussed in this
article. These involve matching techniques (see Ref. \cite{jW01} for a
review) or avoiding introducing an artificial outer boundary
altogether by compactifying spatial infinity
\cite{mClLiOrPfPhV03,fP05}, or making use of hyperboloidal slices and
compactifying null infinity (see, for instance,
\cite{hF81,jF98,sHcStVaZ05}).

\section{Absorbing boundary conditions for the wave equation}
\label{Sect:ABCWaveEq}

To illustrate some of the ideas involved in constructing absorbing
boundary conditions, we start with three examples in order of
increasing difficulty. The simplest example is the flat wave equation
on an interval $(-1,1)$. In this case, perfectly absorbing boundary
conditions can be constructed since the most general solution is
simply the superposition of an arbitrary function of retarded time
with an arbitrary function of advanced time. The second example
discusses the flat wave equation on a three-dimensional ball $B_R$ of
radius $R > 0$. This example is already much more difficult than the
previous one since waves can now travel in infinitely many
directions. One strategy here is to obtain a hierarchy of absorbing
boundary conditions with increasing order of accuracy. The last
example generalizes the second example to a weakly curved
background. When the curvature of the background is taken into
account, a wave solution propagating on Minkowski spacetime acquires
two types of correction terms. The first is a curvature correction
term which obeys Huygens' principle and the second is a fast decaying
term which violates Huygens' principle and describes the backscatter
off curvature. It is important to understand these effects before
constructing absorbing boundaries. An accurate boundary condition must
not eliminate the backscatter which is a real, physical effect! This
example shows that an absorbing boundary condition (as defined in the
introduction) does not mean that {\em all} reflections off the
boundary are to be eliminated, but only {\em spurious} reflections.

\subsection{The flat wave equation on an interval}

Consider the one-dimensional flat wave equation
\begin{equation}
\left( \partial_t^2 - \partial_x^2 \right) u(t,x) = 0, 
\qquad t > 0,\;  x \in [-1,1].
\label{Eq:1DWave}
\end{equation}
The general solution is a superposition of a left- and a right-moving
solution,
\begin{displaymath}
u(t,x) = f_{\nwarrow}(x+t) + f_{\nearrow}(x-t).
\end{displaymath}
Therefore, the boundary conditions
\begin{equation}
(\partial_t - \partial_x) u(t,-1) = 0, \qquad
(\partial_t + \partial_x) u(t,+1) = 0, \qquad t > 0,
\label{Eq:BC1DWave}
\end{equation}
are perfectly absorbing according to our terminology. Indeed, the
operator $b_1 := \partial_t + \partial_x$ has as its kernel the
right-moving solutions $f_{\nearrow}(x-t)$, hence the boundary
condition $b_1 u(t,1) = 0$, $t > 0$, does not ``touch'' these
solutions. On the other hand, $b_1 f_{\nwarrow}(t+x) =
2f'_{\nwarrow}(t+x)$, so the boundary condition at $x=1$ requires that
$f_{\nwarrow}(v) = f_{\nwarrow}(1)$ is constant for advanced time $v =
t+x > 1$. A similar argument shows that the left boundary condition
implies that $f_{\nearrow}(-u) = f_{\nearrow}(-1)$ is constant for
retarded time $u = t - x > 1$. Furthermore, it is known that the
conditions (\ref{Eq:BC1DWave}) together with Eq. (\ref{Eq:1DWave}) and
suitable initial conditions for $u$ and $\partial_t u$ at $t=0$ yield
a well posed IBVP. In particular, the solution is identically zero
after one crossing time ($t \geq 2$) if the initial data has compact
support.

\subsection{The flat wave equation on a three-dimensional ball}

Next, consider the three-dimensional flat wave equation
\begin{equation}
\left( \partial_t^2 - \partial_x^2 - \partial_y^2 - \partial_z^2 \right) u(t,x)
 = 0, \qquad t > 0,\;  (x,y,z) \in B_R,
\label{Eq:3DWave}
\end{equation}
on a ball $B_R$ of radius $R > 0$. The general solution can be
decomposed into spherical harmonics
\begin{displaymath}
u(t,r,\vartheta,\varphi)
 = \frac{1}{r}\sum\limits_{\ell=0}^\infty\sum\limits_{m = -\ell}^{\ell}
    u_{\ell m}(t,r) Y^{\ell m}(\vartheta,\varphi)
\end{displaymath}
which yields the family of reduced equations
\begin{equation}
\left[ \partial_t^2 - \partial_r^2 + \frac{\ell(\ell+1)}{r^2} \right]
u_{\ell m}(t,r) = 0, \qquad t > 0,\; r \in (0,R).
\label{Eq:3DFlatWaveReduced}
\end{equation}
For $\ell=0$ this equation reduces to the previous example, and the
general solution is $u_{00}(t,r) = U_{00\nearrow}(r - t) +
U_{00\nwarrow}(r + t)$ with $U_{00\nearrow}$ and $U_{00\nwarrow}$ two
arbitrary functions. Therefore, the boundary condition
\begin{equation}
{\cal B}_0: \qquad \left. b(r u) \right|_{r=R} = 0,
\end{equation}
where $b := r^2(\partial_t + \partial_r)$ is perfectly absorbing for
spherical waves. For $\ell\geq 1$, exact solutions can be generated
from the solutions for $\ell=0$ by applying suitable differential
operators to $u_{00}(t,r)$. For this, we define the operators
\cite{wB71}
\begin{displaymath}
a_\ell \equiv \partial_r + \frac{\ell}{r}\, ,\qquad
a_\ell^\dagger \equiv -\partial_r + \frac{\ell}{r}
\end{displaymath}
which satisfy the operator identities
\begin{displaymath}
a_{\ell+1} a_{\ell+1}^\dagger = a_\ell^\dagger a_\ell 
 = -\partial_r^2 + \frac{\ell(\ell+1)}{r^2}\; .
\end{displaymath}
As a consequence, for each $\ell=1,2,3...$, we have
\begin{eqnarray}
\left[ \partial_t^2 - \partial_r^2 + \frac{\ell(\ell+1)}{r^2} \right]
a_\ell^\dagger a_{\ell-1}^\dagger ... a_1^\dagger
 &=& \left[ \partial_t^2 + a_{\ell}^\dagger a_{\ell} \right]
   a_\ell^\dagger a_{\ell-1}^\dagger ... a_1^\dagger
\nonumber\\
 &=& a_\ell^\dagger \left[ \partial_t^2 + a_{\ell-1}^\dagger a_{\ell-1} \right]
     a_{\ell-1}^\dagger ... a_1^\dagger
\nonumber\\
 &=& a_\ell^\dagger a_{\ell-1}^\dagger ... a_1^\dagger
     \left[ \partial_t^2 - \partial_r^2 \right].
\nonumber
\end{eqnarray}
Therefore, we have the explicit in- and outgoing solutions
\begin{eqnarray}
u_{\ell m\nwarrow}(t,r)
 &=& a_\ell^\dagger a_{\ell-1}^\dagger ... a_1^\dagger V_{\ell m}(r+t)
  = \sum\limits_{j=0}^\ell (-1)^j\frac{(2\ell-j)!}{(\ell-j)!\, j!} 
   (2r)^{j-\ell} V_{\ell m}^{(j)}(r+t),
\nonumber\\
u_{\ell m\nearrow}(t,r)
 &=& a_\ell^\dagger a_{\ell-1}^\dagger ... a_1^\dagger U_{\ell m}(r-t)
  = \sum\limits_{j=0}^\ell (-1)^j\frac{(2\ell-j)!}{(\ell-j)!\, j!} 
   (2r)^{j-\ell} U_{\ell m}^{(j)}(r-t),
\label{Eq:OutgoingFlatSolution}
\end{eqnarray}
where $V_{\ell m}$ and $U_{\ell m}$ are arbitrary smooth functions
with $j$'th derivatives $V_{\ell m}^{(j)}$ and $U_{\ell m}^{(j)}$,
respectively. In order to construct boundary conditions which are
perfectly absorbing for $u_{\ell m}$, one first shows the following
identity. Let $b = r^2(\partial_t + \partial_r)$ as above, then
\begin{equation}
b^{\ell+1} a_\ell^\dagger a_{\ell-1}^\dagger ... a_1^\dagger 
U(r-t) = 0
\end{equation}
for all $\ell=0,1,2,...$ and all sufficiently smooth functions
$U$. This identity follows easily from
Eq. (\ref{Eq:OutgoingFlatSolution}) and the fact that
$b^{\ell+1}(r^{k}) = k(k+1)\cdot\cdot\cdot (k+\ell) r^{k+\ell+1} = 0$
if $k\in \{ 0,-1,-2,...,-\ell \}$. Therefore, given $L\in \{
1,2,3,...\}$, the boundary condition
\begin{equation}
{\cal B}_L: \qquad \left. b^{L+1}(r u) \right|_{r=R} = 0,
\label{Eq:BCBL}
\end{equation}
leaves the outgoing solutions with $\ell \leq L$ unaltered. Notice
that this condition is {\em local} in the sense that its formulation
does not require the decomposition of $u$ into spherical harmonics.
Furthermore, it was shown in \cite{aBeT80} for domains which can be
more general than $B_R$ that each condition ${\cal B}_L$ yields a well
posed IBVP. By uniqueness this implies that initial data corresponding
to a purely outgoing solution with $\ell \leq L$ yields a purely
outgoing solution (without reflections). In this sense, the condition
${\cal B}_L$ is {\em perfectly absorbing for waves with $\ell \leq
L$}. For waves with $\ell > L$, one obtains spurious reflections;
however, for monochromatic radiation with wave number $k$, the
corresponding amplitude reflection coefficients can be calculated to
decay as $(k R)^{-2(L+1)}$ in the wave zone $k R \gg 1$
\cite{lBoS06}. Furthermore, in most scenarios with smooth solutions,
the amplitudes corresponding to the lower few $\ell$'s will dominate
over the ones with high $\ell$ so that reflections from high $\ell$'s
are unimportant. For a numerical implementation of the boundary
condition ${\cal B}_2$ via spectral methods and a possible application
to General Relativity see Ref. \cite{jNsB04}.

\subsection{The wave equation on a weakly curved background}
\label{Sect:WEQCurved}

Next, we generalize the previous example to the wave equation on a
weakly curved background. More specifically, we consider the wave
equation on the far field region of an asymptotically flat
spacetime. Such a spacetime has the form of Minkowski plus $1/r$
correction terms, where $r$ is the areal radius. The monopolar
correction is given by the $M/r$ term of the Schwarzschild metric,
where $M$ represents the total (Arnowitt-Deser-Misner) mass of the
spacetime. Therefore, to first approximation, we may assume that
spacetime is described by the exterior of a Schwarzschild metric of
mass $M$. In outgoing Eddington-Finkelstein coordinates $(t-r,r)$ the
reduced wave equation on a Schwarzschild background is
\begin{equation}
\left[ \partial_t^2 - \partial_r^2 + \frac{\ell(\ell+1)}{r^2} \right]
u_{\ell m} = -\frac{2M}{r}\left[ (\partial_t + \partial_r)^2
 - \frac{1}{r}(\partial_t + \partial_r) + \frac{1 + \sigma}{r^2}
\right] u_{\ell m}\, .
\label{Eq:3DCurvedWaveReduced}
\end{equation}
Here, $\sigma$ is a parameter which in the present case is zero but is
left arbitrary for future convenience. For $M = 0$, this equation
reduces to the flat reduced wave equation
(\ref{Eq:3DFlatWaveReduced}), for which outgoing solutions have the
form
\begin{displaymath}
u^{(0)}_{\ell m\nearrow}(t,r)
 = a_\ell^\dagger a_{\ell-1}^\dagger ... a_1^\dagger U_{\ell m}(r-t)
\end{displaymath}
with a sufficiently smooth function $U_{\ell m}$. For the following,
we also assume that $U_{\ell m}$ is bounded and vanishes for
sufficiently negative values of its argument. If $M > 0$, we seek an
approximate outgoing solution in the far field region $r \approx R \gg
M$ of the form
\begin{equation}
u_{\ell m\nearrow}(t,r) = u^{(0)}_{\ell m\nearrow}(t,r) 
 + \frac{2M}{R} u^{(1)}_{\ell m\nearrow}(t,r) 
 + O\left( \frac{2M}{R} \right)^2.
\end{equation}
Plugging this into Eq. (\ref{Eq:3DCurvedWaveReduced}) and expanding in
$M/R$, we find that $u^{(1)}_{\ell m\nearrow}$ must satisfy
\begin{eqnarray}
\left[ \partial_t^2 - \partial_r^2 + \frac{\ell(\ell+1)}{r^2} \right]
u^{(1)}_{\ell m\nearrow} 
 &=& -\frac{R}{r}\left[ (\partial_t + \partial_r)^2
     - \frac{1}{r}(\partial_t + \partial_r) + \frac{1 + \sigma}{r^2}
     \right] u^{(0)}_{\ell m\nearrow}
\nonumber\\
 &=& -\frac{R}{r^3}\sum\limits_{j=0}^\ell (-1)^j 
  \frac{(2\ell-j)!}{(\ell-j)! j!}\left[ (\ell+1-j)^2 + \sigma \right]
  (2r)^{j-\ell} U_{\ell m}^{(j)}(r-t).
\label{Eq:FirstOrderCorrectionEq}
\end{eqnarray}
The solution to this equation can be written as the sum of two terms,
\begin{equation}
u^{(1)}_{\ell m\nearrow} (t,r)
 = u^{(curv)}_{\ell m}(t,r) + u^{(backscatter)}_{\ell m}(t,r).
\label{Eq:FirstOrderCorrectionOut}
\end{equation}
The first term includes corrections from the curvature and obeys
Huygens' principle. It has the form
\begin{displaymath}
u^{(curv)}_{\ell m}(t,r)
 = R\sum\limits_{j=0}^{\ell} (-1)^j\frac{(2\ell-j)!}{(\ell-j)! j!}
   (2r)^{j-\ell} c_j U_{\ell m}^{(j+1)}(r-t),
\end{displaymath}
where the coefficients $c_0$, $c_1$, ..., $c_{\ell}$ can be computed
from the recursion relation
\begin{eqnarray}
&& c_\ell = 0, \qquad c_{\ell-1} = 0, \qquad
c_{\ell-2} = -\frac{2(1+\sigma)}{(\ell-1)\ell(\ell+1)(\ell+2)}\; ,
\nonumber\\
&& c_{j-1} = c_j 
 - \frac{2(\ell-j)}{2\ell-j}\;\frac{(\ell-j)^2 + \sigma}{j(j+1)(2\ell+1-j)}\; ,
\qquad j=\ell-2,\ell-3,... 1.
\label{Eq:cjRecursion}
\end{eqnarray}
The second term is a fast decaying term which violates Huygens'
principle and describes the backscatter off the curvature of the
background. It has the form
\begin{equation}
u^{(backscatter)}_{\ell m}(t,r)
 = R(1-\delta_\ell)\frac{(2\ell)!}{(\ell+1)!}
   \frac{\ell^2 + \sigma}{(2r)^{\ell+1}} U_{\ell m}(r-t)
 + R\left( 1 + \delta_\ell\sigma \right)
   \int\limits_{r-t}^\infty K_{\ell}(t,r,x) U_{\ell m}(x) dx,
\label{Eq:Backscatter}
\end{equation}
where $\delta_\ell = 1$ for $\ell=0$ and $\delta_\ell = 0$ for
$\ell\geq 1$, and the integral kernel $K_\ell(t,r,x)$ is given by
\begin{displaymath}
K_\ell(t,r,x) = a_\ell^\dagger a_{\ell-1}^\dagger ... a_1^\dagger
 \frac{1}{(t+r+x)^2}
 = \frac{1}{(2r)^{2+\ell}} \sum\limits_{j=0}^\ell \frac{(2\ell-j)!}{(\ell-j)!} 
   (j+1) \left. z^{-2-j} \right|_{z = \frac{t+r+x}{2r}}\; .
\end{displaymath}
It is not difficult to verify that the expression given in
Eq. (\ref{Eq:FirstOrderCorrectionOut}) indeed solves
Eq. (\ref{Eq:FirstOrderCorrectionEq}) if one notes the recursion
relation (\ref{Eq:cjRecursion}) and the following properties
\begin{displaymath}
\left[ \partial_t^2 - \partial_r^2 + \frac{\ell(\ell+1)}{r^2} \right]
K_\ell(t,r,x) = 0,
\qquad
r^2(\partial_t + \partial_r) K_\ell(t,r,r-t) 
 = -\frac{(2\ell+1)!}{\ell!}\frac{1}{(2r)^{\ell+1}}\; ,
\end{displaymath}
of the integral kernel. Of course, the expression given in
Eq. (\ref{Eq:FirstOrderCorrectionOut}) is not the unique solution to
Eq. (\ref{Eq:FirstOrderCorrectionEq}); one can add an arbitrary
homogeneous solution to it. However, the solution is uniquely
characterized by the following conditions:
\begin{eqnarray}
\lim\limits_{r\to\infty} u^{(1)}_{\ell m\nearrow}(const.+r,r) = 0 &&
\hbox{($u^{(1)}_{\ell m\nearrow}$ vanishes at future null infinity)},
\nonumber\\
\lim\limits_{r\to\infty} u^{(1)}_{\ell m\nearrow}(const.-r,r) = 0 &&
\hbox{($u^{(1)}_{\ell m\nearrow}$ vanishes at past null infinity)},
\nonumber\\
\lim\limits_{t\to\infty} u^{(1)}_{\ell m\nearrow}(t,const.) = 0 &&
\hbox{($u^{(1)}_{\ell m\nearrow}$ vanishes at future time-like infinity)}.
\nonumber
\end{eqnarray}
Notice that for $\ell\geq 1$, the third requirement is necessary in
order to exclude homogeneous solutions of the form $u^{(1)}_{\ell
m\nearrow}(t,r) = (c_1 t + c_0)r^{-\ell}$ (where $c_0$ and $c_1$ are
some constants), which vanish at both future and past null infinity.

Summarizing, outgoing wave solutions have the form
\begin{equation}
u_{\ell m\nearrow}(t,r) 
 = \sum\limits_{j=0}^{\ell} (-1)^j\frac{(2\ell-j)!}{(\ell-j)! j!}
   (2r)^{j-\ell}
   \left[ U_{\ell m}^{(j)}(r-t) + 2M c_j U_{\ell m}^{(j+1)}(r-t)\right]
 + \frac{2M}{R} u^{(backscatter)}_{\ell m}(t,r)
 + O\left( \frac{2M}{R} \right)^2,
\label{Eq:ApproxOutgoing}
\end{equation}
where the coefficients $c_j$ are given in Eq. (\ref{Eq:cjRecursion})
and $u^{(backscatter)}_{\ell m}$ is given in
Eq. (\ref{Eq:Backscatter}). A systematic derivation which includes the
correction terms in $M/R$ of arbitrarily high order is given in
\cite{jBwP73}.

Now let us analyze how much spurious reflection is introduced from
these outgoing wave solutions when the condition ${\cal B}_L$ is
imposed at the boundary surface. If for the moment we neglect
correction terms arising from the backscatter as well as terms which
are quadratic or higher order in $M/R$, then by the same arguments as
in the previous subsection we conclude that ${\cal B}_L$ is perfectly
absorbing for outgoing waves with angular momentum number $\ell$
smaller or equal than $L$. Therefore, ${\cal B}_L$ automatically takes
care of the curvature correction terms $u_{\ell m}^{(curv)}$. If
effects from backscatter are taken into account, ${\cal B}_L$ is not
perfectly absorbing for waves with $\ell\leq L$ anymore, however, in
this case spurious reflections off the boundary surface are very
small. In order to quantify this statement, consider for instance
monopolar scalar radiation ($\ell=0$, $\sigma=0$) for which
\begin{equation}
u_{00\nearrow}(t,r) = U_{00}(r-t) 
 + 2M\int\limits_{r-t}^\infty \frac{U_{00}(x) dx}{(t+r+x)^2}
 + O\left( \frac{2M}{R} \right)^2.
\label{Eq:OutgoingMonopolar}
\end{equation}
Since
\begin{displaymath}
(\partial_t + \partial_r)u_{00\nearrow}(t,r) 
 = -8M\int\limits_{r-t}^\infty \frac{U_{00}(x) dx}{(t+r+x)^3}
 + O\left( \frac{2M}{R} \right)^2
 = -\frac{2M}{r^2}\int\limits_0^\infty \frac{U_{00}(r - t + 2ry)dy}{(1+y)^3}
 + O\left( \frac{2M}{R} \right)^2,
\end{displaymath}
the boundary condition ${\cal B}_0$ is not perfectly absorbing unless
$M=0$. As a consequence, the solution of
Eq. (\ref{Eq:3DCurvedWaveReduced}) consists of a superposition of an
in- and an outgoing wave:
\begin{displaymath}
u_{00}(t,r) = u_{00\nearrow}(t,r) + u_{00\nwarrow}(t,r),
\end{displaymath}
where the ingoing wave has the form $u_{00\nwarrow}(t,r) = V_{00}(r+t)
+ O(2M/R)$ with $V_{00}$ a smooth function. Consider monochromatic
waves of the form
\begin{equation}
U_{00}(r-t) = e^{ik(t-r)}, \qquad
V_{00}(r+t) = \gamma\, e^{-ik(r+t)},
\label{Eq:Monochromatic}
\end{equation}
where $k$ is a given wave number and $\gamma$ is an amplitude
reflection coefficient. For the following, we assume that $0 < k \ll
M^{-1}$. If $k M$ is of the order of unity or larger, powers of $kR$
which are multiplied by $(2M/R)^2$ might be comparable in size or
larger than terms of the form $M/R$ times unity, and in this case the
$(2M/R)^2$-correction terms in $u_{00\nearrow}(t,r)$ might actually be
larger than the $(2M/R)$-correction term. Imposing the boundary
condition ${\cal B}_0$ at $r=R$, the ansatz (\ref{Eq:Monochromatic})
yields
\begin{displaymath}
\gamma = -\frac{M}{R}\frac{e^{2ikR}}{i kR}
 \int\limits_0^\infty \frac{e^{2ikR y} dy}{(1+y)^3}
 + O\left( \frac{2M}{R} \right)^2.
\end{displaymath}
It can be shown that the integral decays as $(k R)^{-1}$ for large $k
R$. Therefore, under the assumption that $M \ll k^{-1} \ll R$ we find
that $|\gamma|$ decays as $(M/R) (kR)^{-2}$. Using similar arguments
it can be shown that the boundary condition ${\cal B}_L$ yields a
reflection coefficient that decays as $(M/R) (kR)^{-(L+2)}$ or faster
for monochromatic waves satisfying $\ell \leq L$ and $M \ll k^{-1} \ll
R$.

Finally, we remark that it is, in principle, possible to improve the
boundary conditions ${\cal B}_L$ in order to take into account the
first order correction terms in $M/R$ of the backscatter. However, by
the very nature of the backscatter, such boundary conditions cannot be
local anymore. As an example, consider again monopolar scalar
radiation where outgoing solutions have the form
(\ref{Eq:OutgoingMonopolar}). Then, the boundary condition
\cite{lBoS07}
\begin{equation}
(\partial_t + \partial_r) u_{00}(t,R)
 + \frac{2M}{R^2}\int\limits_0^{t/2R} \frac{u_{00}(t-2Ry,R) dy}{(1+y)^3} 
 = G(t), \qquad t \geq 0
\label{Eq:BCBackScatter}
\end{equation}
with the boundary data
\begin{displaymath}
G(t) = -\frac{2M}{R^2}\int\limits_{t/2R}^\infty 
   \frac{u_{00}(0,R - t + 2Ry) dy}{(1+y)^3}\; ,
\end{displaymath}
is perfectly absorbing up to (and including) order $2M/R$. Notice that
the integral on the left-hand side of Eq. (\ref{Eq:BCBackScatter})
only involves the past portion $\{ (\tau,R) : 0 \leq \tau \leq t \}$
of the boundary which is available from the past history of a Cauchy
evolution starting at $t=0$. The boundary data $G$ involves an
integral over the initial data at $t=0$ over the region $r > R$
exterior to the computational domain and takes care of the backscatter
that occurred in the past $t < 0$. If the initial data is compactly
supported in the interval $(0,R)$ this integral is zero and can be
discarded. It was shown in \cite{lBoS07} that the boundary condition
(\ref{Eq:BCBackScatter}) is stable in the sense that it admits an
energy estimate. This construction can be repeated for waves with
arbitrary angular momentum number $\ell$.

A different method for constructing absorbing boundary conditions for
linearized gravitational waves propagating on a Schwarzschild
background has recently been presented in \cite{sL04a,sL04b,sL05}.
This method is based on fast converging series expansions of an {\em
exact nonlocal} boundary condition and takes into account arbitrarily
high correction terms in $M/R$ of the Schwarzschild metric. However,
there is no advantage to obtaining a boundary condition which takes
into account the exact form of the Schwarzschild metric beyond the
order of $M/R$ in the construction of boundary conditions for wave
propagation on a asymptotically flat curved background. The reason for
this is that a generic, asymptotically flat background only agrees
with the Schwarzschild metric up to order $M/R$. If second order
effects are to be taken into account, quadratic terms in $M/R$ {\em
and} linear terms in $J/R^2$ (where $J$ is the total angular momentum
of the background) from the background metric must be considered.

\section{Absorbing boundary conditions for Einstein's field equations}
\label{Sect:ABCEinsteinEq}

The construction of absorbing outer boundary conditions in General
Relativity is much more difficult than for the wave equation on a {\em
fixed background} discussed in the previous section. At least three
additional complications arise. First, in the Cauchy problem of
General Relativity, Einstein's field equations split into a set of
evolution equations and a set of constraints. If the spatial time
slices are infinite or compact without boundaries it can be shown via
the use of Bianchi's identities that any smooth enough solution of the
evolution equations with constraint satisfying initial data
automatically satisfies the constraints everywhere and at all
times. However, if the time slices possess a nonempty boundary, this
statement holds only if constraint-preserving boundary conditions are
specified. The second complication is due to the fact that
gravitational waves do not propagate on a fixed background but deform
the spacetime metric as they evolve. As a consequence, it is not clear
how to ``fix'' the boundary geometrically. It would be nice if one
could specify the boundary conditions in such a way that the embedding
of the boundary surface in the resulting spacetime is independent of
the coordinate choice for which the evolution is performed. Otherwise,
two evolutions using different coordinates might obtain different
portions of spacetime even if they both start with the same initial
slice and data. Finally, the third complication stems from the
nonlinear nature of Einstein's field equations. In particular, the
superposition principle for wave propagation does not hold so it is
much harder to superpose an outgoing and and ingoing wave as was done
in the previous section in order to quantify the amount of spurious
reflection.

\subsection{Constraint-preserving boundary conditions}

The construction of constraint-preserving boundary conditions is
probably the best understood and most studied issue of all the
complications listed above: see Refs.
\cite{jS98,hFgN99,bSbSjW02,bSjW03,gCjPoRoSmT03,gCoS03,gC-PhD-03,sFrG03a,sFrG03b,sFrG04a,sFrG04b,cGjM04a,cGjM04b,oRoS05,oSmT05,lKlLmSlBhP05,nT-PhD-05,lLmSlKrOoR06,gNoS06,hKjW06,oR06,dAnT06,aA06}
for analytic studies and Refs.
\cite{mIoR02,gClLmT02,jBlB02,bSbSjW02,bSjW03,
gC-PhD-03,lLmSlKhPdSsT04,mHlLrOhPmSlK04,oSmT05,lKlLmSlBhP05,cBtLcPmZ05,oR-PhD-05,mBbSjW06,mBhKjW06,oRlLmS07}
for numerical studies. The basic idea in constructing
constraint-preserving boundary conditions is to derive the constraint
propagation system, describing the propagation of constraint
violations, and to impose boundary conditions for this system which
ensure that zero is the only solution with trivial initial data. As an
example, consider Einstein's field equations in harmonic coordinates,
\begin{displaymath}
\Gamma^a := \Box_g x^a = 0,
\end{displaymath}
where $\Box_g$ denotes the d'Alembertian operator with respect to the
metric $g_{ab}$. In these coordinates, Einstein's vacuum equations
reduce to a set of ten coupled, quasilinear wave equations of the form
\begin{displaymath}
g^{cd}\partial_c\partial_d g_{ab} = F_{ab}(g)[\partial g,\partial g],
\end{displaymath}
subject to the constraint $\Gamma^a = 0$, where $F_{ab}(g)[\partial
g,\partial g]$ depends quadratically on the first derivatives of the
metric fields $g_{ab}$. Using the twice contracted Bianchi identities,
one finds that as a consequence of the evolution equations, $\Gamma^c$
satisfies a wave equation on its own,
\begin{displaymath}
g^{cd}\partial_c\partial_d\Gamma^a = L^a(g,\partial g,\partial^2 g)[\Gamma]
\end{displaymath}
where $L^a(g,\partial g,\partial^2 g)[\Gamma]$ is a {\em linear first
order differential operator} in $\Gamma^a$ with coefficients depending
on $g_{ab}$ and its partial derivatives up to second order. Therefore,
the initial conditions $\left. \Gamma^a \right|_{t=0} = 0$,
$\left. \partial_t\Gamma^a \right|_{t=0} = 0$ insure that $\Gamma^a =
0$ on the domain of dependence of the initial slice. If the initial
slice is a Cauchy slice, this implies that $\Gamma^a \equiv 0$
everywhere on the spacetime, but if time-like boundaries are present,
one needs to impose additional conditions at the boundary in order to
guarantee $\Gamma^a = 0$ everywhere. There are different ways of
assuring that $\Gamma^a \equiv 0$ is the only solution with trivial
initial data $\left. \Gamma^a \right|_{t=0} = 0$,
$\left. \partial_t\Gamma^a \right|_{t=0} = 0$. The simplest way is to
specify Dirichlet data \cite{hKjW06}
\begin{displaymath}
\left. \Gamma^a \right|_{\partial\Omega} = 0.
\end{displaymath}
However, other conditions, such as Sommerfeld-type conditions
\cite{oR06,lLmSlKrOoR06,mShPlLlKoRsT06,oRlLmS07} or higher-order
absorbing boundary conditions \cite{mRoRoS07} are also possible.

\subsection{Gauge-controlling boundary conditions}

In the above example of Einstein's equations in harmonic coordinates,
where the evolution equations have the form of ten coupled wave
equations, one needs ten boundary conditions. As illustrated above,
four conditions are needed in order to insure constraint
propagation. Since there are two gravitational degrees of freedom, one
expects that two boundary conditions are needed in order to control
gravitational radiation. The remaining four boundary conditions are
related to the residual freedom in choosing harmonic coordinates and
fixing the geometry of the boundary surface. Preliminary ideas about
how to specify such ``gauge'' controlling boundary conditions are
given in \cite{oRlLmS07,mRoRoS07}, but it is not clear yet how these
conditions can be used in order to fix (at least part of) the geometry
of the boundary surface. An exception are the boundary conditions
constructed in \cite{hFgN99} for a tetrad formulation of Einstein's
vacuum equations, which specify the mean curvature of the boundary
surface as embedded in spacetime to be an arbitrary constant.

Perhaps even more important than fixing the geometry of the outer
boundary is the ability to specify a unique radial coordinate $r$ and
a unique outward radial null vector $l^a$ at each point of the
boundary. Such quantities are needed for the generalization of the
hierarchy of boundary conditions ${\cal B}_L$ defined in
Eq. (\ref{Eq:BCBL}), where $b = r^2 l^a\nabla_a$. A recent proposal
for constructing $r$ and $l^a$ based on the assumption that near the
boundary, spacetime can be represented as Schwarzschild plus a small
perturbation thereof is given in \cite{lBoS07}. However, it is not yet
completely clear how to identify the Schwarzschild background in this
proposal.

\subsection{Absorbing wave boundary conditions}

Once constraint- and gauge-controlling boundary conditions have been
specified, the next step is to construct boundary conditions which
control the physical degrees of freedom by minimizing the amount of
spurious gravitational reflection off the boundary surface. If the
boundary is placed far from the strong field region, the field
equations can be linearized about a weakly curved spacetime near the
outer boundary. As discussed in Sect. \ref{Sect:WEQCurved}, it is
sufficient to consider a Schwarzschild background provided $R \gg M$
where $R$ is the radius of the outer boundary and $M$ the total
(Arnowitt-Deser-Misner) mass of the system. Therefore, near the outer
boundary, it is safe to assume that spacetime can be written as
Schwarzschild plus a small perturbation thereof.

Linear perturbations on a Schwarzschild background can be described by
the Regge-Wheeler-Zerilli formalism \cite{tRjW57,fZ70,oSmT01}. By
performing a decomposition of the metric perturbations into spherical
tensor harmonics, one obtains in this formalism two families of master
equations for gauge-invariant potentials $\Phi_{\ell m}^{(\pm)}$
describing even $(+)$ and odd $(-)$ parity metric fluctuations with
angular momentum numbers $\ell\geq 2$ and $|m| \leq \ell$. In
particular, the metric perturbations in the Regge-Wheeler gauge can be
reconstructed from the potentials $\Phi_{\ell m}^{(\pm)}$ {\em without
solving additional differential equations} \cite{oSmT01}. To first
order in $M/R$, the master equations for $\Phi_{\ell m}^{(\pm)}$ have
the form (\ref{Eq:3DCurvedWaveReduced}), where $\sigma^{(+)} = -4 +
3/[(\ell-1)(\ell+2)]$ in the even-parity case and $\sigma^{(-)} = -4$
in the odd-parity case. Therefore, approximate outgoing solutions have
the form (\ref{Eq:ApproxOutgoing}) and in principle, the boundary
conditions ${\cal B}_L$: $\left. b^{L+1}\Phi_{\ell m}^{(\pm)}
\right|_{r=R} = 0$, where $b = r^2(\partial_t + \partial_r)$, can be
applied to the gauge-invariant quantities $\Phi_{\ell m}^{(\pm)}$.
However, the relation between $\Phi_{\ell m}^{(\pm)}$ and the metric
perturbations is nonlocal in the sense that it depends on the angular
momentum number $\ell$. Therefore, applying ${\cal B}_L$ in this way
results in a nonlocal boundary conditions and its implementation
requires a decomposition into spherical harmonics at the boundary. An
alternative way is to first compute the (linearized) Weyl scalar
$\Psi_0$ from $\Phi_{\ell m}^{(\pm)}$ which is also a gauge-invariant
quantity and to formulate the boundary condition on $\Psi_0$. If
$(t-r,r)$ denote outgoing Eddington-Finkelstein coordinates for the
Schwarzschild background, and $\Psi_0 = C_{abcd} l^a m^b l^c m^d$ is
constructed from the Weyl tensor $C_{abcd}$ and the null vectors
$l^a\partial_a = \partial_t + \partial_r$, $m^a\partial_a = (\sqrt{2}
r)^{-1}\left( \partial_\vartheta +
i\sin^{-1}\vartheta\;\partial_\varphi \right)$ the boundary conditions
\cite{jB07a,lBoS07}\footnote{The normalization of the null vector
$l^a$ differs from the one chosen in \cite{lBoS07} by a factor of
$\sqrt{2/N}$, where $N = 1 - 2M/r$. This explains the absence of the
factor $N^{-1}$ in Eq. (\ref{Eq:BCCL}).}
\begin{equation}
{\cal C}_L: \qquad \left. 
\partial_t\left[ r^2\left( \partial_t + \partial_r \right) \right]^{L-1}
\left( r^5\Psi_0 \right) \right|_{r=R} = 0,\qquad
L = 1,2,3,...
\label{Eq:BCCL}
\end{equation}
are perfectly absorbing for linearized gravitational waves with
angular momentum number $\ell \leq L$ to first order in $M/R$ if
backscatter is neglected. Here, the time derivative operator
$\partial_t$ in front of the operator inside the square brackets is
introduced in order to allow for a static contribution to $\Psi_0$.
Notice that for $L=1$ this condition just freezes $\Psi_0$ to its
initial value. This freezing-$\Psi_0$ boundary condition has been
given before in formulations of the IBVP of Einstein's equations
\cite{hFgN99,jBlB02,oSmT05,oR06,gNoS06}. (Actually, the formulations
in Refs. \cite{hFgN99,oSmT05,gNoS06} also consider more general
boundary conditions which allow to couple $\Psi_4$ to $\Psi_0$ but it
is unclear if this coupling is useful for reducing spurious
reflections.) In this sense, the hierarchy ${\cal C}_L$ of boundary
conditions improves the freezing-$\Psi_0$ one. Reflection coefficients
due to spurious reflections from backscatter for quadrupolar waves are
computed in \cite{lBoS07}. It is found that for a spherical outer
boundary and quadrupolar gravitational radiation, ${\cal C}_2$ reduces
spurious reflections by a factor of $(15M/2R)(kR)^{-2}$ compared to
the freezing-$\Psi_0$ condition ${\cal C}_1$ when $k R > 1$. In
\cite{lBoS07}, a new boundary condition ${\cal D}_2$ similar to the
the condition (\ref{Eq:BCBackScatter}) is derived which takes into
account first order correction terms in $M/R$ of the backscatter for
quadrupolar linear waves. It would be interesting to generalize this
analysis to take into account second order effects. However, in this
case, it is not sufficient to consider perturbations of a
Schwarzschild background since quadratic effects in $M/R$ {\em and}
linear effects in $J/R^2$ (where $J$ is the total angular momentum of
the spacetime) from the full metric should also be included.

\subsection{Well posedness results}

Once constraint-preserving absorbing boundary conditions with the
desired properties have been specified, the next step is to prove the
well posedness of the resulting IBVP. That is, one has to prove that
for given initial data $u_0$ in an appropriate function space there
exists a unique solution $u(t)$ of the evolution equations in a time
interval $[0,T]$ which satisfies the constructed boundary conditions
and such that $u(0) = u_0$. Furthermore, one needs to show that $u(t)$
depends continuously on the initial and boundary data in the sense
that if $u^{(n)} \to u_0$ is a sequence of initial data converging to
$u_0$ then $u^{(n)}(t)$ converges to $u(t)$ for each $t\in
[0,T]$. This property is important for the convergence of a numerical
approximation since in this case the initial data always contains
errors. Finally, one has to check that if $u_0$ satisfies the
constraint equation, so does $u(t)$ for each $t\in [0,T]$.

A well posed IBVP for Einstein's vacuum equations was presented in
Ref. \cite{hFgN99}. This work, which is based on a tetrad formulation,
recasts the evolution equations into first order symmetric hyperbolic
quasilinear form with maximally dissipative boundary conditions
\cite{kF58,pLrP60} for which (local in time) well posedness is
guaranteed \cite{pS96b}. There has been considerable effort to obtain
well posed formulations for the more commonly used metric formulations
of gravity. Partial results using similar mathematical techniques as
in \cite{hFgN99} were obtained in
\cite{bSbSjW02,bSjW03,gCjPoRoSmT03,cGjM04a,cGjM04b,nT-PhD-05,dAnT06,aA06}.
However, most of these works are either restricted to the linearized
equations or to reflecting (and not absorbing) boundary conditions.
For results on coupled hyperbolic-elliptic linear problems with
constraint-preserving boundary conditions based on semigroup
techniques see \cite{oRoS05,gNoS06}.

A different technique for showing the well posedness of the IBVP is
based on the frozen coefficient principle where one freezes the
coefficients of the evolution and boundary operators. In this way, the
problem is simplified to a linear, constant coefficient problem on the
half-space which can be solved explicitly by using a Fourier-Laplace
transformation \cite{KL89}. This method yields a simple algebraic
condition (the determinant condition) which is necessary for the well
posedness of the IBVP. Work based on verifying the determinant
condition for the Einstein case is given in
\cite{jS98,gCoS03,cGjM04a,cGjM04b,oSmT05,oR06}. Sufficient conditions
for the well posedness of the frozen coefficient problem were
developed by Kreiss \cite{hK70}. Kreiss' theorem provides a stronger
form of the determinant condition whose satisfaction leads to well
posedness if the evolution system is strictly hyperbolic. One of the
key results in \cite{hK70} is the construction of a smooth symmetrizer
for the problem for which well posedness can be shown via an energy
estimate in the frequency domain. Using the theory of
pseudo-differential operators it is expected that the verification of
Kreiss' condition also leads to well posedness for quasilinear
problems, like Einstein's field equations. Work based on the
verification of Kreiss' condition in the Einstein case is given in
\cite{jS98,oR06} but since in that case the evolution system is not
strictly hyperbolic it is not clear if these results imply well
posedness. Recently, Kreiss and Winicour \cite{hKjW06} introduced a
new pseudo-differential first order reduction of the wave equation
which leads to a strictly hyperbolic system. Using this reduction they
were able to prove well posedness of the IBVP for Einstein's field
equations in harmonic coordinates in the frozen coefficient
approximation. This work is generalized to higher-order absorbing
boundary conditions in \cite{mRoRoS07}. For an alternative proof of
the results in \cite{hKjW06} which does not require a
pseudo-differential first order reduction see \cite{hKoRoSjW07}.

\subsection{Applications to numerical relativity}

For applications of constraint-preserving boundary conditions to
numerical relativity, see Refs. \cite{mIoR02,gClLmT02} for simulations
of self-gravitating scalar fields in spherical symmetry,
Ref. \cite{jBlB02} for the simulation of 1D colliding gravitational
plane waves, Ref. \cite{oR-PhD-05} for evolutions of Brill waves in
axisymmetry,
Refs. \cite{bSbSjW02,bSjW03,gC-PhD-03,cBtLcPmZ05,lKlLmSlBhP05,oSmT05,mBbSjW06,mBhKjW06,lLmSlKrOoR06,oRlLmS07}
for tests in three spatial dimensions, Ref. \cite{mShPlLlKoRsT06} for
binary black hole simulations and Refs. \cite{oSlL04,fGlLoS07} for the
simulation of bubble spacetimes in five-dimensional theories of
gravity. In particular, Refs. \cite{lKlLmSlBhP05,oSmT05} implement a
first order symmetric hyperbolic formulation of Einstein's vacuum
equations with the freezing-$\Psi_0$ boundary condition, and
Refs. \cite{lLmSlKrOoR06,mShPlLlKoRsT06,oRlLmS07} also freeze $\Psi_0$
at the boundary but use the harmonic formulation of the field
equations. In \cite{oRlLmS07}, constraint-preserving freezing-$\Psi_0$
boundary conditions are tested for the case of a perturbed
Schwarzschild black hole and compared to other types of boundary
conditions proposed in the literature. It is found that the version of
constraint-preserving freezing-$\Psi_0$ boundary conditions in
\cite{oRlLmS07} performs better than all alternate boundary treatments
tested. It should be interesting to numerically implement the boundary
conditions ${\cal C}_L$, $L \geq 2$, given in Eq. (\ref{Eq:BCCL})
which are refinements of the freezing-$\Psi_0$ boundary condition.

\section{Conclusions}
\label{Sect:Conclusions}

Formulating absorbing outer boundary conditions for the numerical
solution of Einstein's field equations involves five steps: i) The
construction of constraint-preserving boundary conditions which make
sure that no constraint-violating modes enter the computational
domain, ii) finding boundary conditions that geometrically control the
evolution of the boundary surface, iii) finding conditions that
minimize the amount of spurious reflection of gravitational radiation
off the boundary, iv) proving well posedness of the resulting
initial-boundary value problem (IBVP) and v) discretizing the problem.

As discussed in this article, there has been a lot of effort in
carrying out step i) which is, by now, well-understood. In contrast to
this, step ii) needs further work. Regarding step iii), a promising
approach for minimizing spurious reflections is the hierarchy ${\cal
C}_L$ of local boundary conditions on the Weyl scalar $\Psi_0$
presented in \cite{lBoS07}. They have the property of being perfectly
absorbing including curvature corrections (but neglecting backscatter)
to order $M/R$ for all multipoles of gravitational radiation up to
$L$, where $M$ is the Arnowitt-Deser-Misner mass of the spacetime and
$R$ a typical radius of the boundary surface. However, their precise
formulation requires a radial coordinate and an outward radial null
vector field at the boundary whose unambiguous definition is an open
problem and could benefit from progress in step ii). Regarding step
iv), a complete proof of the well posedness of the IBVP has been given
in \cite{hFgN99} for a tetrad formulation of Einstein's field
equations. Recently, proofs for well posedness have also been given
for the frozen coefficient limit of Einstein's equations in harmonic
coordinates \cite{hKjW06,mRoRoS07} and it is expected that these
results can be generalized to the full Einstein equations based on the
theory of pseudo-differential operators. Finally, in step v),
promising results have been achieved in the numerical implementation
of the harmonic formulation with constraint-preserving absorbing
boundary conditions \cite{mBhKjW06,lLmSlKrOoR06,mShPlLlKoRsT06}. It is
expected that the boundary condition ${\cal C}_2$ proposed in
\cite{lBoS07}, which is perfectly absorbing for quadrupolar linearized
gravitational radiation, will improve these results. Finally, it would
be interesting to develop similar boundary conditions for the
Baumgarte-Shapiro-Shibata-Nakamura \cite{mStN95,tBsS99} formulation of
Einstein's field equations which is often used in numerical
relativity. For partial results along these lines, see
\cite{hBoS04,cGjM04b,sFrG04b}.

\begin{acknowledgments}
It is a pleasure to thank J. Bardeen, L. Buchman, L. Lehner, G. Nagy,
O. Reula, O. Rinne, M. Tiglio and J. Winicour for many enlightening
discussions during my work on boundary conditions. The author also
thanks L. Buchman for reading the manuscript and helpful
suggestions. This work was partially supported by grant CIC-4.20 to
Universidad Michoacana.
\end{acknowledgments}

\bibliography{refs}
\end{document}